\documentclass[12pt]{article}

\usepackage[centertags]{amsmath}
\usepackage{amsmath}
\usepackage{amsfonts}
\usepackage{amssymb}
\usepackage{amsthm}
\usepackage{newlfont}
\usepackage{amscd}
\usepackage{empheq}
\usepackage{epsfig,graphics,graphicx,amsfonts,psfrag}
\usepackage{verbatim}
\usepackage{eucal}

\newcommand{\NN}{{\mathbb N}}

\newcommand{\RR}{{\mathbb R}}
\newcommand{\ZZ}{{\mathbb Z}}

\newcommand{\CC}{{\mathbb C}}
\newcommand{\beq}{\begin{equation}}
\newcommand{\eeq}{\end{equation}}
\newcommand{\ba}{\begin{array}}
\newcommand{\ea}{\end{array}}
\newcommand{\bea}{\begin{eqnarray}}
\newcommand{\eea}{\end{eqnarray}}
\newcommand{\eps}{{\epsilon}}

\usepackage{graphicx}
\usepackage[usenames,dvipsnames]{color}
\usepackage{transparent}
\usepackage{float}

\begin{document}
\begin{center}
{\bf The periodic $N$ breather anomalous wave solution\\ of the Davey-Stewartson equations;\\ first appearance, recurrence, and blow up properties}
\vskip 15pt
{\it F. Coppini $^{1,3}$, P. G. Grinevich $^{2,4}$ and
  P. M. Santini $^{1,5}$}

\vskip 20pt

{\it
$^1$ Dipartimento di Fisica, Universit\`a di Roma "La Sapienza", and \\
Istituto Nazionale di Fisica Nucleare (INFN), Sezione di Roma, \\ 
Piazz.le Aldo Moro 2, I-00185 Roma, Italy

\smallskip
 
$^2$ Steklov Mathematical Institute of Russian Academy of Sciences,\\ 8 Gubkina St., Moscow, 199911, Russia.
}

\vskip 10pt
$^{3}$e-mail:  {\tt francesco.coppini@uniroma1.it,\\ francesco.coppini@roma1.infn.it}\\
$^{4}$e-mail:  {\tt pgg@landau.ac.ru}\\
$^{5}$e-mail:  {\tt paolomaria.santini@uniroma1.it \\ paolo.santini@roma1.infn.it}\\
\bigskip
\vskip 10pt

{\today}

\end{center}

\begin{abstract}
The integrable focusing Davey-Stewarson (DS) equations, multidimensional generalizations of the focusing cubic nonlinear Schr\"odinger (NLS) equation, provide ideal mathematical models for describing analytically the dynamics of $2+1$ dimensional anomalous (rogue) waves (AWs). In this paper i) we construct the $N$-breather AW solution of Akhmediev type of the DS1 and DS2 equations, describing the nonlinear interaction of $N$ unstable modes over the constant background solution. ii) For the simplest multidimensional solution of DS2 we construct its limiting subcases, and we identify the constraint on its arbitrary parameters giving rise to blow up at finite time. iii) We use matched asymptotic expansions to describe the relevance of the constructed AW solutions in the doubly periodic Cauchy problem for small initial perturbations of the background, in the case of one and two unstable modes. We also show, in the case of two unstable modes, that no blow up takes place generically, although the AW amplitude can be arbitrarily large. All the results are expressed in terms of elementary functions.
\end{abstract}

\section{Introduction}

Davey-Stewartson (DS) type equations \cite{DS} describe the amplitude modulation of weakly nonlinear quasi monochromatic waves in $2+1$ dimensions, and are relevant in nonlinear optics, water waves, plasma physics and Bose condensates \cite{Benney,DS,ABB,Nishinari,Huang}. Only a sub class of these equations are integrable, for special choices of their constant parameters, and can be written in the form:
\beq\label{DS}
\ba{l}
i u_t+u_{xx}-\nu u_{yy}+2\eta q u=0, \ \ \eta=\pm 1, \ \ \nu=\pm 1, \\
q_{xx}+\nu q_{yy}= (|u|^2)_{xx}-\nu (|u|^2)_{yy}, \\
x,y,t\in\RR^3, \ u=u(x,y,t)\in\CC, \ q=q(x,y,t)\in\RR,
\ea
\eeq
where $u$ is the complex amplitude of the monochromatic wave, and the real field $q(x,y,t)$ is related to the mean flow. If $\nu=-1$ we have the DS1 equation (surface tension prevails on gravity in the water wave derivation); in this case the sign of $\eta$ is irrelevant, since one can go from the equation with $\eta=-1$ to the equation with $\eta=1$ via the changes $q\to -q$ and $x\leftrightarrow y$; therefore there exists only one DS1 equation \cite{GS5}. If $\nu=1$, gravity prevails on surface tension and we have the DS2 equations; in this case the sign of $\eta$ cannot be rescaled away and we distinguish between focusing and defocusing DS2 equations for respectively $\eta=1$ and $\eta=-1$. It turns out that the shallow water limit of the Benney-Roskes equations \cite{Benney} leads to the DS1 and to the defocusing DS2 equations \cite{AS}, and DS1 plays a relevant role in the description of the initial-boundary value problem for dromions \cite{Fokas1,Fokas2}, exponentially localized solutions first discovered via B\"acklund transformations \cite{BLMP}. We also remark that DS2 plays a relevant role in the theory of immersion of surfaces in $\RR^4$ \cite{Konopel1,PedPink,Konopel2,Taim1,Taim2,Taim6}. The DS equations \eqref{DS} are integrable $2+1$ dimensional generalizations of the celebrated nonlinear Schr\"odinger (NLS) equations
\beq\label{NLS}
i v_t+v_{xx}+2\eta |v|^2 v=0, \ \ \eta=\pm 1,\ \ x,t\in\RR, \ v=v(x,t)\in\CC,
\eeq
reducing to them when there is no $y$ dependence. The focusing ($\eta=1$) NLS equation is the simplest nonlinear integrable model describing modulation instability (MI), and MI is considered the main physical mechanism for the creation of anomalous waves (AWs) in nature \cite{HendersonPeregrine,Dysthe,Osborne,KharifPeli1,KharifPeli2,Onorato2}.

Concerning the NLS Cauchy problem for initial perturbations of the unstable background, what we call the Cauchy problem for AWs, if such a perturbation is localized, then slowly modulated periodic oscillations described by the elliptic solution of (\ref{NLS}) play a relevant role in the longtime regime \cite{Biondini1,Biondini2}. Using the finite-gap method, the NLS periodic Cauchy problem of AWs was recently solved to leading order \cite{GS1,GS3} in the case of a finite number of unstable modes, leading to a quantitative description of the recurrence properties of the dynamics in terms of the multi-breather generalization \cite{ItsRybinSall} of the Akhmediev breather (AB) solution \cite{Akhmed0}
\beq\label{def_Akhmed}
\ba{l}
Akh(x,t,\phi):=e^{2it}{\cal A}(x,t,\phi), \\
{\cal A}(x,t,\phi):=\frac{\cosh[2\sin(2\phi)t+2i\phi]-\sin(\phi)\cos[2\cos(\phi)x]}{\cosh[2\sin(2\phi)t]+\sin(\phi)\cos[2\cos(\phi)x]},
\ea
\eeq
where $\phi$ is an arbitrary real parameter. In the simplest case of one unstable mode only, this theory describes quantitatively a Fermi-Pasta-Ulam-Tsingou (FPUT) recurrence of AWs described by the AB \eqref{def_Akhmed} \cite{GS1,GS2}. In addition, a finite-gap perturbation theory for 1+1 dimensional AWs has been also developed \cite{Coppini1}, see also \cite{Coppini2,Coppini3}, to describe analytically the order one effects of physical perturbations of the NLS model on the AW dynamics. See also \cite{GS2} for an alternative approach to the study of the AW recurrence, based on matched asymptotic expansions. See \cite{GS6} for the study of the instability properties of the AB solution within the NLS dynamics, and \cite{GS4} for a finite-gap model describing the numerical instabilities of the AB. See \cite{GS7} for the analytic study of the phase resonances in the AW recurrence. See \cite{San}, \cite{CS_AL1,CS_AL2}, and \cite{CS_MTM} for the analytic study of the AW recurrence in other NLS type models: respectively the nonlocal PT-symmetric NLS equation \cite{AM1}, the discrete Ablowitz-Ladik model \cite{AL}, and the relativistic Massive Thirring model \cite{Thirring}, showing analytically the universal features of the AW recurrence in the periodic setting. MI and AWs of integrable multicomponent NLS equations have also been investigated \cite{BCDegaLombOnorWab,BDegaCW,DegaLombSommacal,DegaLombSommacal2}. We also remark that the NLS recurrence of AWs in the periodic setting has been investigated in several numerical and real experiments, see, f.i., \cite{Kimmoun,Yuen1,Yuen2,Mussot,Pierangeli}, and qualitatively studied in the past via a 3-wave approximation of NLS \cite{Infeld,Trillo}.

As it was discussed in \cite{GS5}, the integrable focusing DS2 equation \eqref{DS} ($\nu=\eta=1$) is the best mathematical model on which to construct an analytic theory of $2+1$ dimensional AWs, and a finite gap formalism allowing one to solve in principle, to leading order, the doubly periodic Cauchy problem for AWs of the focusing DS2 equation has been recently constructed \cite{GS5}.

Although the physical relevance of DS2 is not clear at the moment, a $2+1$ dimensional generalization of the AW perturbation theory developed in \cite{Coppini1,Coppini2,Coppini3} could be used in principle to treat non integrable physically relevant multidimensional NLS models with mean flow as perturbations of the integrable DS equations. This will be the subject of future investigation. 

We remark that the homogeneous background solution $u_0=a e^{2i\eta |a|^2 t},\ \ q_0=|a|^2$ of equations \eqref{DS}, where $a$ is an arbitrary complex parameter, can be simplified to
\beq\label{background}
u_0(x,y,t)=1, \ \ q_0(x,y,t)=0,
\eeq
using the scaling symmetry and the gauge symmetry
\begin{equation}
 \label{eq:DS_gauge} 
u(x,y,t) \rightarrow u(x,y,t) \exp\left(-i\frac{\eta}{2}\int^t f(\tau) d\tau   \right), \ \ q (x,y,t) \rightarrow  q(x,y,t) + f(t)
\end{equation}
of \eqref{DS} \cite{GS5}, where $f(t)\in\RR$ is an arbitrary function of time, and in the rest of the paper we use such a background.

Some exact AW solutions  of the DS equations are already known in the literature (see for example \cite{Liu1,Liu2,Otha1, Otha2}) and, in contrast with the focusing NLS equation, DS2 solutions corresponding to smooth Cauchy data may blow up at finite time \cite{Ozawa,Otha2,KleinSaut0,KleinSaut,Taim9,Taim10,TaimTz}.

In this paper i) we construct the $N$-breather AW solution of Akhmediev type of equations \eqref{DS}, describing the nonlinear interaction of $N$ unstable modes over the constant background solution \eqref{background}. ii) We select, in the DS2 case, the subclass of AW solutions that are relevant in the Cauchy problem for periodic AWs, and, in the case of the simplest multidimensional solution, we identify the constraint on its arbitrary parameters giving rise to blow up at finite time. iii) We construct limiting cases of this  multidimensional solution. iv) We use the matched asymptotic expansions technique introduced in \cite{GS2} to describe the relevance of the constructed AW solutions in the DS2 doubly periodic Cauchy problem for AWs, in the case of one and two unstable modes, showing in particular that blow up is not generic.

For the description of the AW dynamics when the number of unstable modes is higher, matched asymptotic expansion techniques are not adequate, as in the NLS case, and must replaced by a suitable implementation of the finite gap formalism developed in \cite{GS5}; this will be the subject of a subsequent paper.

\begin{figure}[H]
	\centering
        \includegraphics[width=8.7cm]{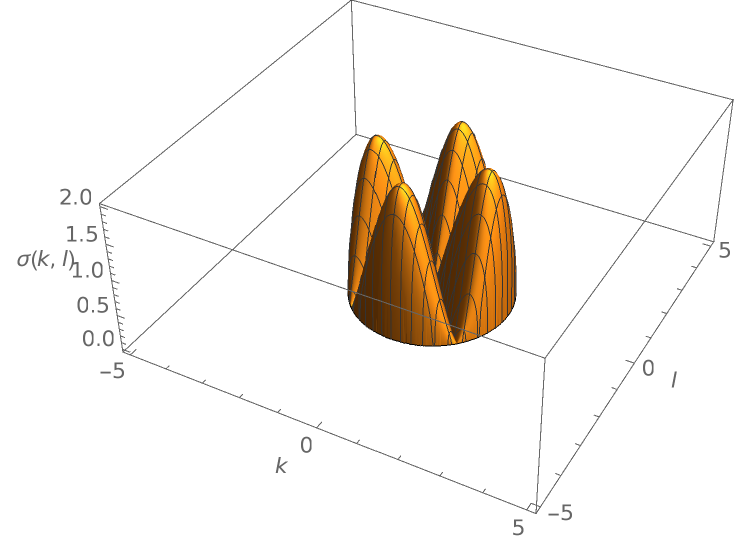}
	\caption{For the focusing DS2 equation, the growth rate $\sigma(k,l)$ in the instability region $k^2+l^2<4$.}
	\label{instDS2a}
\end{figure}

To study the modulation instability properties of the background solution \eqref{background}, we slightly perturb it as follows
\beq
u=1+\eps(f+ig), \ \  q=\eps w, \ \ \eps\ll 1, \ f,g,w\in\RR.
\eeq
Then $f,g,w$ satisfy the linear PDEs
\beq\label{linearized}
\ba{l}
f_t+g_{xx}-\nu g_{yy}=0, \ \ g_t-f_{xx}+\nu f_{yy}-2\eta w=0, \\
w_{xx}+\nu w_{yy}=2(f_{xx}-\nu f_{yy}).
\ea
\eeq
Looking for a  monochromatic perturbation
\beq
\ba{l}
f=Ue^{i(kx+ly)+\sigma t}+cc, \ g=Ve^{i(kx+ly)+\sigma t}+cc, \\
w=We^{i(kx+ly)+\sigma t}+cc, \ \ k,l\in\RR ,
\ea
\eeq
one obtains the following system of homogeneous equations 
\beq
\begin{pmatrix}
  \sigma & -(k^2-\nu l^2) & 0 \\
  k^2-\nu l^2 & \sigma & -2\eta \\
  -2(k^2-\nu l^2) & 0 & k^2+\nu l^2
\end{pmatrix}\begin{pmatrix}U\\V\\W\end{pmatrix}=0 ,
\eeq
and the condition for the existence of nontrivial solutions gives
\beq\label{growth_rate_DS2}
\sigma^2(k,l)=\frac{(k^2-\nu l^2)^2[4\eta -(k^2+\nu l^2)]}{k^2+\nu l^2}.
\eeq
Therefore we have the following stability properties of the background \eqref{background}.\\
For DS1 ($\nu=-1,\eta=1$): $\sigma^2(k,l)=\frac{(k^2+l^2)^2[4 -(k^2-l^2)]}{k^2-l^2}$. If $k^2-l^2>4$, then $\sigma^2<0$, $\sigma\in i\RR$, and the background is neutrally stable. If $0<k^2-l^2<4$, then $\sigma^2>0$ and the background is unstable with growth rate
\beq\label{sigma_DS1}
\sigma(k,l)=\frac{(k^2+l^2)\sqrt{4-(k^2-l^2)}}{\sqrt{k^2-l^2}}.
\eeq
For DS2 ($\nu=1$): $\sigma^2(k,l)=\frac{(k^2-l^2)^2[4\eta -(k^2+l^2)]}{k^2+l^2}$.  If $\eta=-1$, then $\sigma^2<0$, $\sigma\in i\RR$ and the background is neutrally stable. If $\eta=1$ we have two cases. If $k^2+l^2>4$, then $\sigma^2 <0$ and the bachground is stable. If
\beq\label{instability}
|\vec k|^2=k^2+l^2<4, \ \ \mbox{and} \ \ k^2\ne l^2 , \ \ \vec k =(k,l),
\eeq
then $\sigma^2>0$ and the background is unstable with exponential growth rate
\beq
\sigma(k,l)=|\Omega(k,l)|, \ \ \Omega(k,l)=\frac{(k^2-l^2)\sqrt{4-(k^2+l^2)}}{\sqrt{k^2+l^2}}.
\eeq
Therefore no AWs are associated with the defocusing DS2, while AWs are present in the focusing DS2 equation for sufficiently small wave vectors $\vec k$, in perfect analogy with the NLS case (see Figure \ref{instDS2a}).

We observe that a convenient parametrization of the unstable modes of DS1 and DS2 reads as follows
\beq\label{parametrization}
\ba{l}
\mbox{DS1}: \ k=2\cos(\phi)\cosh(\theta), \ \ {l}=2\cos(\phi)\sinh(\theta), \ \Rightarrow \ \sigma=2\sin(2\phi)\cosh(2\theta), \\ 
\mbox{DS2}: \ k=2\cos(\phi)\cos(\theta), \ \ {l}=2\cos(\phi)\sin(\theta), \ \Rightarrow \ \Omega=2\sin(2\phi)\cos(2\theta). 
\ea
\eeq
with
\beq\label{parametrization}
\ba{l}
\mbox{DS1}: \ \ \ \phi =\arccos\left(\frac{\sqrt{k^2 -l^2}}{2} \right), \ \ \ \theta=\tanh^{-1} \left(\frac{l}{k} \right), \\
\mbox{DS2}: \ \ \ \phi =\arccos\left(\frac{\sqrt{k^2 +l^2}}{2} \right), \ \ \ \theta=\arctan\left(\frac{l}{k} \right).
\ea
\eeq

\section{The $N$-breather quasi homoclinic solutions}

Exact explicit solutions of integrable soliton equations can be constructed through different methods: a suitable genus zero degeneration of the finite gap method \cite{ItsRybinSall,BBEIM,Krichever2,Krichever3}, the Darboux  transformations \cite{Matveev0}, dressing techniques \cite{ZakharovShabatdress,ZakharovManakov,ZakharovMikha}, and the Hirota method \cite{Hirota0,Hirota}. Here we use the Hirota method to constructs the $N$-breather ($N\in\NN^+$) AW solutions of the DS equations \eqref{DS}, the $2+1$ dimensional generalizations of the N-breather solution of Akhmediev type of the focusing NLS equation \cite{ItsRybinSall}, describing the nonlinear interaction of $N$ unstable modes.

The Hirota bilinear form of the DS equations \eqref{DS}, corresponding to solutions over the background \eqref{background}, reads \cite{Boling}:
\beq
\ba{l}
(i D_t+D_x^2 -\nu D_y^2)G\cdot F=0, \\
(D_x^2 +\nu D_y^2)F\cdot F=2\eta (|G|^2-F^2),
\ea
\eeq
where $D_{x}$ is the Hirota derivative with respect to the generic independent variable $x$ \cite{Hirota}, so that
\beq
\ba{l}
D_x G\cdot F=G_x(x)F(x)-G(x)F_x(x), \\
D^2_x G\cdot F=G_{xx}(x)F(x)-2 G_x(x)F_x(x)+G(x)F_{xx}(x),
\ea
\eeq
and functions $G$ and $F$ are related to the solution of equations \eqref{DS} as follows:
\beq\label{def_breather1}
\ba{l}
u(x,y,t)=\frac{G(x,y,t)}{F(x,y,t)}, \ \ F(x,y,t)\in\RR, \ \ G(x,y,t)\in\CC, \\
\ \\
q(x,y,t)=(\partial_x^2-\nu \partial_y^2)\log\left(F(x,y,t)\right);
\ea
\eeq
in addition:
\beq\label{def_breather2}
|u(x,y,t)|^2=1+(\partial_x^2+\nu \partial_y^2)\log\left(F(x,y,t)\right).
\eeq

Then the $N$-breather AW solutions of the DS1 and DS2 equations are described by the following formulas in terms of elementary functions:
\beq\label{def_breather3}
\ba{l}
F(x,y,t)=\!\!\!\!\!\!\!\!\!\sum\limits_{\begin{tiny}\ba{c} n_j=0,1\\ 1\le j\le 2N\ea\end{tiny}}\!\!\!\!\!\!\!\!\exp\!\left(\sum\limits_{j=1}^{2N}n_j\zeta_j(x,y,t)+\!\!\!\!\!\!\sum\limits_{1\le j<k\le 2N}\!\!\!\!\!b_{jk}n_j n_k \right), \\
G(x,y,t)=\!\!\!\!\!\!\!\!\!\!\sum\limits_{\begin{tiny}\ba{c} n_j=0,1\\ 1\le j\le 2N\ea\end{tiny}}\!\!\!\!\!\!\!\!(-1)^{\sum\limits_{j=1}^{2 N}n_j}\!\!\!\exp\!\left(\sum\limits_{j=1}^{2N}n_j(\zeta_j(x,y,t)+2i\hat\phi_j)+\!\!\!\!\!\!\sum\limits_{1\le j<k\le 2N}\!\!\!\!\!b_{jk}n_j n_k \right).
\ea
\eeq
For DS1:
\beq\label{def_breather4_DS1}
\ba{l}
\zeta_j(x,y,t)=\left\{
  \ba{ll}
  i\left[{k}_j x+{l}_j y+\zeta_{0j}\right]+\Omega_j t, & 1\le j\le N, \\
  -i\left[{k}_{j-N} x+{l}_{j-N} y+\zeta_{0j}\right]+\Omega_{j-N} t, & N+1\le j\le 2N,
  \ea\right. \\
{k}_j=2\cos(\phi_j)\cosh(\theta_j), \\
{l}_j=2\cos(\phi_j)\sinh(\theta_j), \\
\Omega_j=\frac{{k}_j^2+{l}_j^2}{\sqrt{{k}_j^2 -{l}_j^2}}\sqrt{4-\left({k}_j^2 -{l}_j^2\right)}=2\sin(2\phi_j)\cosh(2\theta_j), 
\ea
\eeq
\beq\label{def_breather5_DS1}
\ba{l}
b_{jk}=\left\{\ba{ll}
  \log\left(\frac{\cosh\left(\hat\theta_j -\hat\theta_k\right)-\cos(\hat\phi_j -\hat\phi_k)}{\cosh\left(\hat\theta_j -\hat\theta_k\right)+\cos\left(\hat\phi_j +\hat\phi_k\right)}\right), & 1\le j<k\le N, \ N\!+\!1\!\le\! j\!<\!k\!\le \!2 N, \\
  \log\left(\frac{\cosh\left(\hat\theta_j -\hat\theta_k\right)+\cos\left(\hat\phi_j -\hat\phi_k\right)}{\cosh\left[\nu\left(\hat\theta_j -\hat\theta_k\right)\right]-\cos\left(\hat\phi_j +\hat\phi_k\right)}\right), & 1\le j\le N \mbox{ and } N+1\le k\le 2 N,  
  \ea\right.\\
\ \\
\hat\phi_j=\left\{
  \ba{ll}
\phi_j, &  1\le j\le N, \\
\phi_{j-N}, &  N+1\le j\le 2N.
  \ea\right.
\ea
\eeq
For DS2:
\beq\label{def_breather4_DS2}
\ba{l}
\zeta_j(x,y,t)=\left\{
  \ba{ll}
  i\left[{k}_j x+{l}_j y+\zeta_{0j}\right]+\Omega_j t, & 1\le j\le N, \\
  -i\left[{k}_{j-N} x+{l}_{j-N} y+\zeta_{0j}\right]+\Omega_{j-N} t, & N+1\le j\le 2N,
  \ea\right. \\
{k}_j=2\cos(\phi_j)\cos(\theta_j), \\
{l}_j=2\cos(\phi_j)\sin(\theta_j), \\
\Omega_j=\frac{{k}_j^2-{l}_j^2}{\sqrt{{k}_j^2 +{l}_j^2}}\sqrt{4-\left({k}_j^2 + {l}_j^2\right)}=2\sin(2\phi_j)\cos(2\theta_j), \ \ \sigma_j =|\Omega_j|, 
\ea
\eeq
\beq\label{def_breather5_DS2}
\ba{l}
b_{jk}=\left\{\ba{ll}
  \log\left(\frac{\cos\left(\hat\theta_j -\hat\theta_k\right)-\cos(\hat\phi_j -\hat\phi_k)}{\cos\left(\hat\theta_j -\hat\theta_k\right)+\cos\left(\hat\phi_j +\hat\phi_k\right)}\right), & 1\le j<k\le N, \ N\!+\!1\!\le\! j\!<\!k\!\le \!2 N, \\
 \log\left(\frac{\cosh\left(\hat\theta_j -\hat\theta_k\right)+\cos\left(\hat\phi_j -\hat\phi_k\right)}{\cos\left(\hat\theta_j -\hat\theta_k\right)-\cos\left(\hat\phi_j +\hat\phi_k\right)}\right), & 1\le j\le N \mbox{ and } N+1\le k\le 2 N,  
  \ea\right.\\
\ \\
\hat\phi_j=\left\{
  \ba{ll}
\phi_j, &  1\le j\le N, \\
\phi_{j-N}, &  N+1\le j\le 2N,
\ea\right. ~ 
\hat\theta_j=\left\{
  \ba{ll}
\theta_j, &  1\le j\le N, \\
\theta_{j-N}, &  N+1\le j\le 2N.
  \ea\right.
\ea
\eeq

The proof of this result, in the Hirota method philosophy, is by induction. To the best of our knowledge, only the solution for $N=1$ is known in the literature \cite{Boling}, but this solution describes a straight line (one dimensional) AW that can be easily derived from the AB solution of NLS using elementary symmetries.

\subsection{Doubly periodic AWs}

\begin{figure}[H]
  \centering
  \includegraphics[width=6.75cm]{instDS2a0.png}\includegraphics[width=6.75cm]{instDS2a1.png}
        \includegraphics[width=6.75cm]{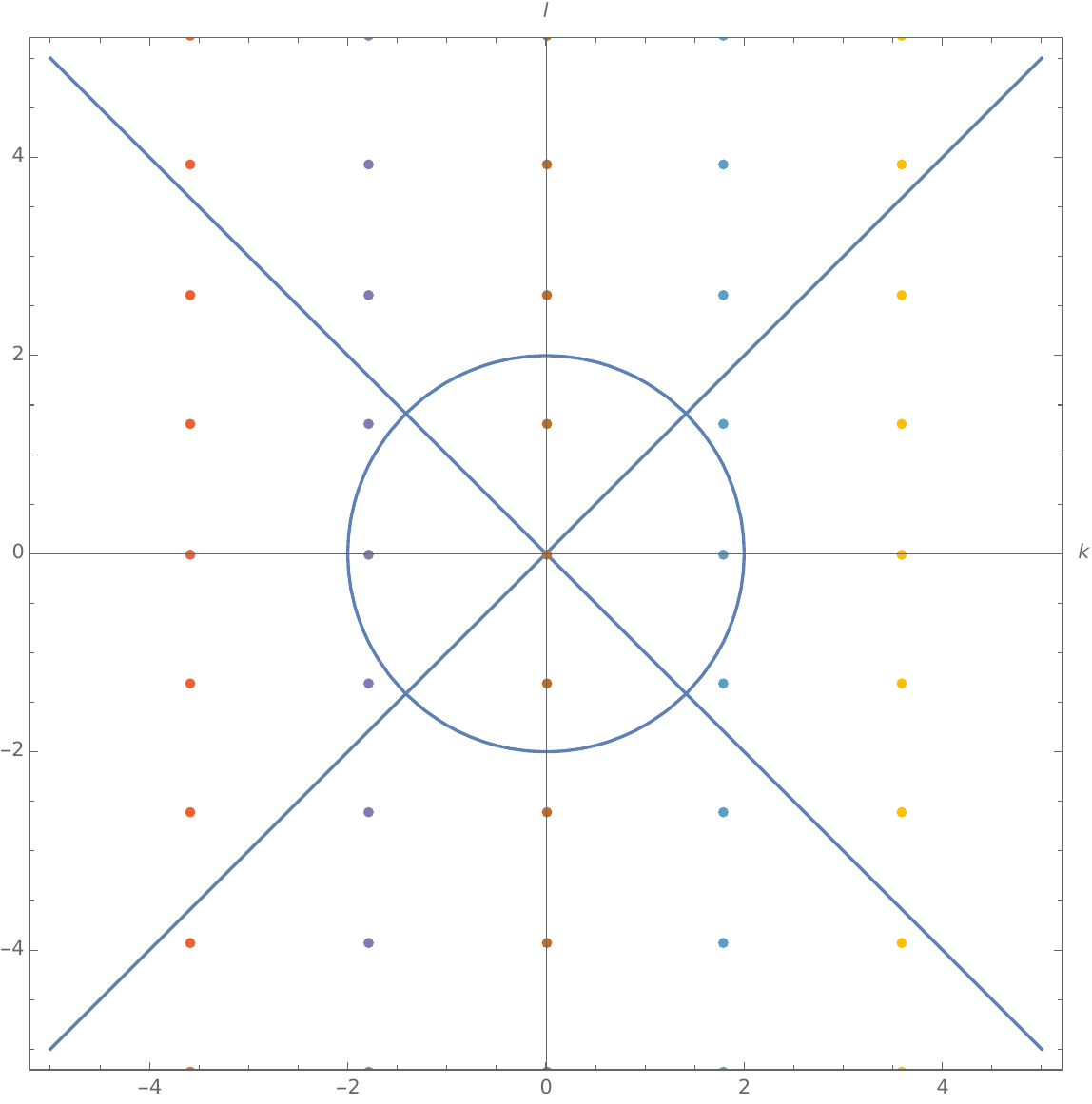}
        \includegraphics[width=6.75cm]{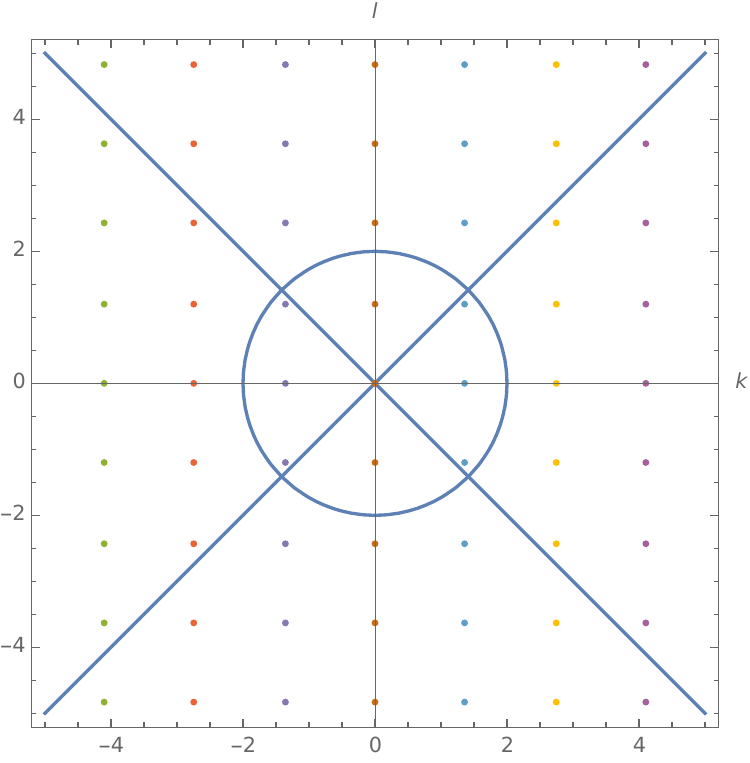}
	\caption{For the DS2 equation, the instability region in the $\vec k=(k,l)$ plane is the disk $k^2+l^2<4$, with $k^2-l^2\ne 0$. The Fourier modes of the linearized theory are $\vec k_{m,n}=2\pi(\frac{m}{L_x},\frac{n}{L_y})$, where $m,n\in\ZZ$, and $L_x$ and $L_y$ are respectively the periods in the $x$ and $y$ directions. In the top left picture $L_x=3.5,~L_y=2.8$, and there is only the unstable mode $\pm\vec k_{1,0}$. In the top right picture $L_x=2.8,~L_y=3.5$, and there is only the unstable mode $\pm\vec k_{0,1}$. In the bottom left picture $L_x=3.5,~L_y=4.8$, and there are only the $2$ unstable modes $\pm\vec k_{1,0},\pm\vec k_{0,1}$. In the bottom right picture  $L_x=4.6,~L_y=5.2$, and there are only the four unstable modes $\pm\vec k_{1,0},\pm\vec k_{0,1},\pm\vec k_{1,1},\pm\vec k_{1,-1}$. }
	\label{instDS2b}
\end{figure}

As it was shown in \cite{GS5}, in the well-posed doubly-periodic DS2 Cauchy problem of AWs with periods $L_x$ and $L_y$, the wave vectors of the above $N$-breather solution are quantized as follows 
\beq\label{quantized_vectors}
\ba{l}
\vec k_{m,n}=(k_m,l_n), \ \ k_m=\frac{2\pi}{L_x}m, \ l_n=\frac{2\pi}{L_y}n, \ \ m,n\in\ZZ, \ \ L_x\ne L_y, 
\ea
\eeq
and lie on the rectangular lattice of Figure \ref{instDS2b}, constrained by the instability condition
\beq\label{constraints_quantized_vectors}
k_m^2 +l_n^2 <4 \ \ \ \Leftrightarrow \ \ \ \left(\frac{m}{L_x} \right)^2+\left(\frac{n}{L_y} \right)^2<\frac{1}{\pi^2}. 
\eeq

The simplest possible instability configurations are, in order of complication, the following.\\
1) The case in which there is only one unstable mode, the mode $\pm\vec k_{1,0} =\pm(k_1,0)$ on the $k$ axis, with:
\beq\label{k10}
1<k_1 <2, \ \ l_1>2 \ \ \ \Leftrightarrow \ \ \ \pi< L_x <2\pi, \ \ L_y <\pi,
\eeq
or the mode $\pm\vec k_{0,1} =\pm(0,l_1)$ on the $l$ axis, with:
\beq\label{k01}
1<l_1 <2, \ \ k_1>2 \ \ \ \Leftrightarrow \ \ \ \pi< L_y <2\pi, \ \ L_x <\pi;
\eeq
see respectively the top left and top right pictures of Figure \ref{instDS2b}.\\
2) The case in which there are only the two unstable modes $\pm\vec k_{1,0},\pm\vec k_{0,1}$, with
\beq\label{k10_k01}
1<k_1,l_1 <2, \ \ k^2_1 +l^2_1 >4  \ \ \ \Leftrightarrow \ \ \ \pi< L_x,L_y <2\pi, \ \ \frac{1}{L^2_x}+\frac{1}{L^2_y}>\frac{1}{\pi^2};
\eeq
see the bottom left picture of  Figure \ref{instDS2b}.\\
3) The case in which there are only the four unstable modes $\pm\vec k_{1,0}$,$\pm\vec k_{0,1}$,$\pm\vec k_{1,1}$,\\$\pm\vec k_{1,-1}$, with
\beq\label{k10_k01_k11_k1-1}
1<k_1,l_1 <2, \ \ k^2_1 +l^2_1 <4  \ \ \ \Leftrightarrow \ \ \ \pi< L_x,L_y <2\pi, \ \ \frac{1}{L^2_x}+\frac{1}{L^2_y}<\frac{1}{\pi^2};
\eeq
see the bottom right picture of  Figure \ref{instDS2b}. Increasing the periods $L_x$ and $L_y$, higher order modes enter the instability region and the picture becomes more and more complicated. In this paper we limit our considerations to the first two cases 1) and 2), postponing to a subsequent paper the study of a higher number of unstable modes.

\subsection{The simplest cases}

\paragraph{N=1.} The solution \eqref{def_breather1}-\eqref{def_breather5_DS1} reads, after some manipulation
\beq\label{u_N=1}
\ba{l}
u_1(x,y,t)=\frac{\cosh\left[\sigma (t-t_1)+2i\phi \right]
-\sin(\phi)\cos\left[k x+l y +\zeta\right]}{\cosh\left[\sigma (t-t_1)\right]
+\sin(\phi)\cos\left[k x+l y +\zeta\right]}, 
\ea
\eeq
\beq\label{q_N=1}
\ba{l}
q_1(x,y,t)=\cos(2\theta)\left(|u_1(x,y,t)|^2 -1 \right),
\ea
\eeq
where
\beq
\sigma =2\sin(2\phi)\cos(2\theta), \ \ k=2\cos\phi \cos\theta, \ \ l=2\cos\phi \cos\theta .
\eeq

This solution describes a straight line breather parallel to the line $x\cos(\theta)+y\sin(\theta)=0$ of arbitrary slope $\cot(\theta)$ (due to the arbitrariness of the parameter $\theta$), repeated periodically in the $x$ and $y$ directions with periods $L_x=2\pi/k$ and $L_y=2\pi/l$, and decaying to the background \eqref{background} at $t\to\pm\infty$.

Since its appearance changes the phase of the background by the factor $4\phi$, it can be viewed as a quasi homoclinic solution as the AB solution \eqref{def_Akhmed} of NLS, and actually can be written in terms of the AB itself as follows:
\beq
u_1(x,y,t)={\cal A}\left(\cos(\theta)(x-x_1)+\sin(\theta) (y-y_1),\cos(2\theta)(t-t_1),\phi\right).
\eeq

In the doubly periodic Cauchy problem of the AWs, only its two limiting cases in which the solution depends on just one space variable play a role (see Figure \ref{instDS2b}). The $y$ dependence disappears if $l=0$ ($\theta=0$), and the DS2 solution \eqref{u_N=1},\eqref{q_N=1} reads:
\beq\label{u10}
u_{1,0}(x,t)={\cal A}(x-x_1,t-t_1,\phi), \ \ q_{1,0}(x,t)=|u_{1,0}(x,t)|^2 -1,
\eeq
describing a straight line breather parallel to the $y$ axis (case 1, top left picture in Figure \ref{instDS2b}). The $x$ dependence disappears if $k=0$ ($\theta=\pi/2$), and the solution \eqref{u_N=1},\eqref{q_N=1} of DS2 reduces to
\beq\label{u01}
\ba{l}
u_{0,1}(y,t)=e^{2it}{\cal A}(y-y_1,-(t-t_1),\phi_1), \ \  q_{0,1}(y,t)=1-|u_{0,1}(y,t)|^2,
\ea
\eeq
describing a straight line breather parallel to the $x$ axis (case 1, top right picture in Figure \ref{instDS2b}).

  As in the NLS case, it is always possible to construct the rational limit of \eqref{u_N=1},\eqref{q_N=1} when $k_1,l_1$ tend to zero (for $\phi_1\to\pm \pi/2$). If $\phi_1\to\pi/2$, then $u_1\to -\exp{2it}$. If  $\phi_1\to\pi/2$, up to an irrelevant minus sign, the breather tends to the following generalization of the Peregrine solution
\beq\label{Peregrine}
\ba{l}
u_{1P}(x,y,t)=1-\frac{4+16\cos(2\theta)(t-t_1)}{1+16 (t-t_1)^2\cos^2(2\theta)+4(\cos(\theta_1)(x-x_1)+\sin(\theta)(y-y_1))^2}, \\
\ \\
q_{1P}(x,y,t)=1+\frac{8\cos(2\theta)\left[1+16\cos^2(2\theta)(t-t_1)^2-4(\cos(\theta)(x-x_1)+\sin(\theta)(y-y_1))^2\right]}{(1+16 (t-t_1)^2\cos^2(2\theta)+4(\cos(\theta)(x-x_1)+\sin(\theta)(y-y_1))^2)^2},
\ea
\eeq
constant on the line $x\cos\theta+y\sin\theta=0$ with arbitrary slope, and rationally localized over the background in any other direction.

\paragraph{N=2.} The straight line breather solution \eqref{u_N=1},\eqref{q_N=1} and its rational limit \eqref{Peregrine} could have been constructed from the AB plus symmetry considerations. The simplest truly two dimensional AW describes the interaction of the  horizontal $\pm\vec k_{1,0}=(k_1,0)$ and vertical $\pm\vec k_{0,1}=(0,l_1)$ unstable modes, where
\beq\label{def_k1_l1}
k_1=\frac{2\pi}{L_x}=2\cos\phi_{1,0}, \ \ l_1=\frac{2\pi}{L_y}=2\cos\phi_{0,1}, \ \ \theta_{1,0}=0, \ \theta_{0,1}=\pi/2,
\eeq
corresponding to the conditions
\beq
\pi<L_x,L_y <2\pi \ \Leftrightarrow \ 1<k_1,l_1<2 \ \Leftrightarrow \ 0<\phi_{1,0},\phi_{0,1}<\pi/3.   
\eeq
Then the solution \eqref{def_breather1}-\eqref{def_breather5_DS1} reads, after some manipulation:
\beq\label{xy_sol}
u_2(x,y,t;\phi_{1,0},\phi_{0,1},x_0,y_0,t_{1,0},t_{0,1},\rho)=\frac{N(x,y,t)}{D(x,y,t)}e^{i\rho},
\eeq
\beq\label{xy_sol_N}
\ba{l}
N(x,y,t)=\cosh\left[\sigma_{1,0}(t-t_{1,0})+\sigma_{0,1}(t-t_{0,1})+2i(\phi_{1,0}-\phi_{0,1}))\right] \\
+b_{12}^2\cosh\left[\sigma_{1,0}(t-t_{1,0})-\sigma_{0,1}(t-t_{0,1})+2i(\phi_{1,0}+\phi_{0,1}))\right] \\
-2 b_{12}\Big(\sin\phi_{1,0}\cos(X_{1,0})\cosh\left[\sigma_{0,1}(t-t_{0,1})-2i\phi_{0,1}\right]\\
+\sin\phi_{0,1}\cos(Y_{0,1})\cosh\left[\sigma_{1,0}(t-t_{1,0})+2i\phi_{1,0}\right]
+\sin\phi_{1,0}\sin\phi_{0,1}\cos(X_{1,0})\cos(Y_{0,1})\Big),
\ea
\eeq
\beq\label{xy_sol_D}
\ba{l}
D(x,y,t)=\cosh\left[\sigma_{1,0}(t-t_{1,0})+\sigma_{0,1}(t-t_{0,1})\right] \\
+b_{12}^2\cosh\left[\sigma_{1,0}(t-t_{1,0})-\sigma_{0,1}(t-t_{0,1}))\right] \\
+2 b_{12}\Big(\sin\phi_{1,0}\cos(X_{1,0})\cosh\left[\sigma_{0,1}(t-t_{0,1})\right]\\
+\sin\phi_{0,1}\cos(Y_{0,1})\cosh\left[\sigma_{1,0}(t-t_{1,0})\right]-\sin\phi_{1,0}\sin\phi_{0,1}\cos(X_{1,0})\cos(Y_{0,1})\Big),
\ea
\eeq
where
\beq\label{xy_sol_param}
\ba{l}
X_{1,0}=k_1 (x-x_0)=2\cos(\phi_{1,0})(x-x_0), \ \ Y_{0,1}=l_1 (y-y_0)=2\cos(\phi_{0,1})(y-y_0), \\
\ \\
\sigma_{1,0}=k_1\sqrt{4-k_1^2}=2\sin(2\phi_{1,0}), \ \ \sigma_{0,1}=l_1\sqrt{4-l_1^2}=2\sin(2\phi_{0,1})=-\Omega_{0,1},\\
\ \\
b_{12}=\frac{\cos(\phi_{1,0}-\phi_{0,1})}{\cos(\phi_{1,0}+\phi_{0,1})},
\ea
\eeq
and $\rho,x_0,y_0,t_{1,0},t_{0,1}$ are arbitrary real parameters.

If, in addition,
\beq\label{stablek11_k1-1}
\frac{1}{L_x^2}+\frac{1}{L_y^2}>\frac{1}{\pi^2} \ \ \Leftrightarrow \ \ k_1^2 +l_1^2>4 \ \ \Leftrightarrow \ \ \cos^2\phi_{1,0}+\cos^2\phi_{0,1}>1,   
\eeq
then the modes $\pm\vec k_{1,1}=(k_1,l_1)$ and $\pm\vec k_{1,-1}=(k_1,-l_1)$ are stable and $b_{12}>0$ (see the bottom left plot of Figure \ref{instDS2b}); instead, if 
\beq\label{unstablek11_k1-1}
\frac{1}{L_x^2}+\frac{1}{L_y^2}<\frac{1}{\pi^2} \ \ \Leftrightarrow \ k_1^2 +l_1^2<4 \ \Leftrightarrow \ \ \cos^2\phi_{1,0}+\cos^2\phi_{0,1}<1,   
\eeq
then also the modes $\pm\vec k_{1,1}=(k_1,l_1)$ and $\pm\vec k_{1,-1}=(k_1,-l_1)$ are unstable and $b_{12}<0$ (see the bottom right plot of Figure \ref{instDS2b}). It is quite clear that the solution \eqref{xy_sol}-\eqref{xy_sol_param} will be relevant in the periodic Cauchy problem for AWs only under the constraint \eqref{stablek11_k1-1}.

While the parameters $\rho,x_0,y_0$ and one of the parameters $t_{1,0},t_{0,1}$ (say, $t_{0,1}$) are associated with space-time translation and elementary gauge symmetries of DS2, the additional parameter $t_{1,0}$ is associated with the integrability of the model. 

\begin{figure}[H]
	\centering
        \includegraphics[width=9.0cm]{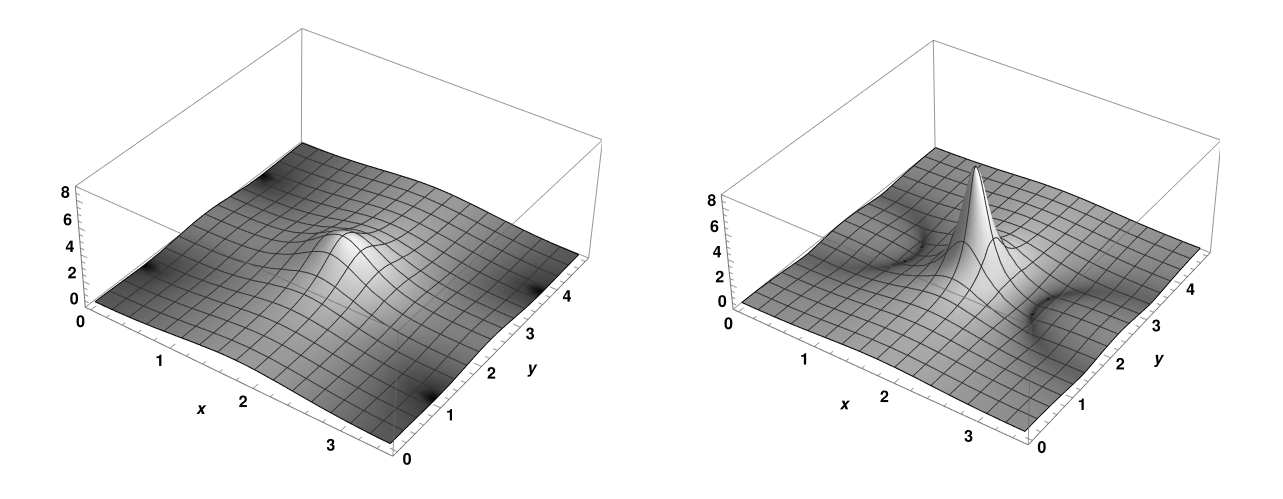} \\
        \includegraphics[width=9.0cm]{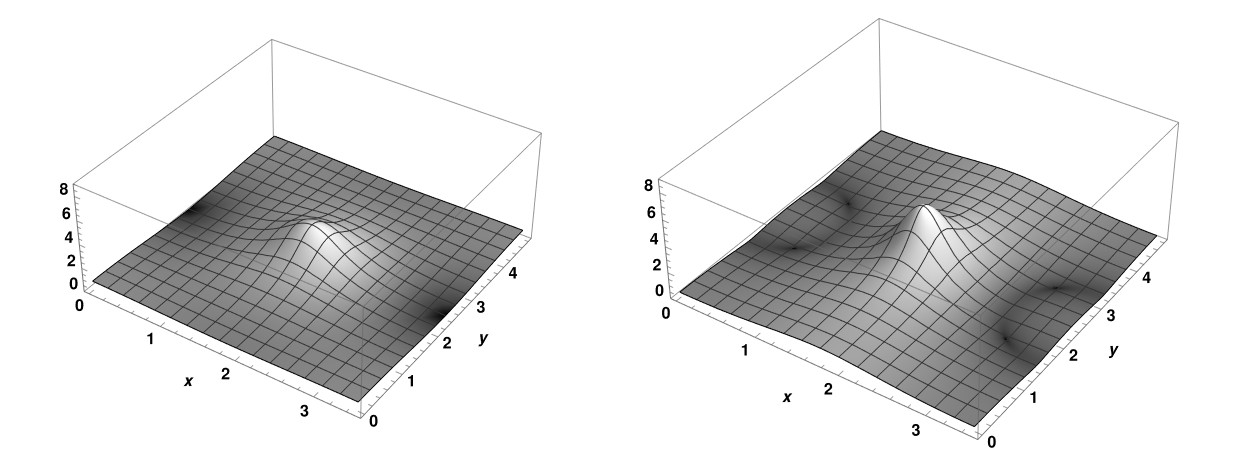}\\
        \includegraphics[width=4cm]{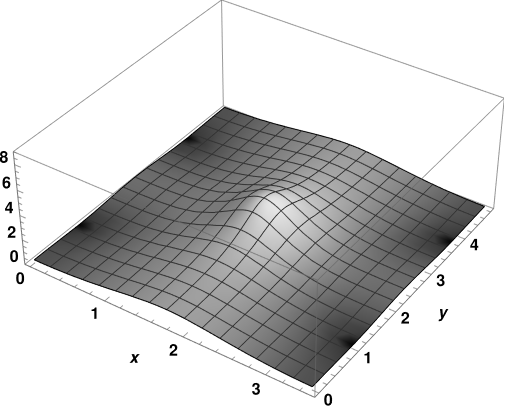}
      \caption{Five snapshots of the evolution of the 2-breather AW solution \eqref{xy_sol} in a basic period ($L_x=3.5$, $L_y=4.8$), describing the nonlinear interaction of two unstable modes, one parallel to the $x$ axis and the other parallel to the $y$ axis with parameters $t_{1,0}=0$ and $t_{0,1}=-0.1$. Top left: the growth of the AW from the background ($t=-1.2$); top right: the first emergence  ($t=-0.83$); medium left: between two emergences ($t=-0.17$); medium right: second emergence ($t=0.78$); bottom: the disappearance of the AW into the background ($t=1.1$). For generic parameters, the solution is smooth and the two emergences, occurring both at $(x,y)=(x_0+L_x/2,y_0+L_y/2)$, are different. In this case $x_0=y_0=0$.}\label{AW2}
\end{figure}

For generic parameters, the solution \eqref{xy_sol}-\eqref{xy_sol_param} decays to the backgrounds $\exp\left[i\left(\rho\pm (\phi_{1,0}-\phi_{0,1}) \right)\right]$ as $t\to\pm\infty$, and describes the nonlinear interaction between the horizontal and vertical unstable modes. Since the associated growth rates $\sigma_{1,0},\sigma_{0,1}$ are generically different, it describes two consecutive appearances in time of $2+1$ dimensional doubly-periodic smooth bumps, both located at $(x_0+L_x/2,y_0+L_y/2)$ (see Figure \ref{AW2}).

There are however two non generic choices of the parameter $t_{10}-t_{01}$:
\beq\label{critical_parameter}
\ba{l}
t_{10}-t_{01}=\pm \Delta_c, \\
\Delta_c:=\frac{1}{\sigma_{1,0}}\log\left(B+\sqrt{B^2-1}\right)-\frac{1}{\sigma_{0,1}}\log\left(A+\sqrt{A^2-1}\right),
\ea
\eeq
where
\beq\label{def_A_B}
\ba{l}
A=\cos\phi_{1,0}\tan(\phi_{1,0}+\phi_{0,1})-\frac{\sin\phi_{0,1}}{\cos(\phi_{1,0}-\phi_{0,1})}, \\
B=\cos\phi_{0,1}\tan(\phi_{1,0}+\phi_{0,1})-\frac{\sin\phi_{1,0}}{\cos(\phi_{1,0}-\phi_{0,1})},
\ea
\eeq
for which the solution \eqref{xy_sol} blows up at the critical times
\beq
t^{\pm}_c=t_{01}\pm\frac{1}{\sigma_{1,0}}\log\left(B+\sqrt{B^2-1}\right).
\eeq
More precisely, if $t_{10}-t_{01}=-\Delta_c$, the solution blows up at its first appearance in $(x_0,y_0,t=t^-_c)$; if $t_{10}-t_{01}=\Delta_c$, the solution blows up at its second appearance in $(x_0,y_0,t=t^+_c)$. The AW solution \eqref{xy_sol} behaves as follows near the blow up:
\beq
\ba{l}
u_2(x,y,t,{t^{\pm}_{10}}_c)\sim \frac{N(x_0,y_0,t^{\pm}_c)e^{i\rho}}{a(t-t^{\pm}_c)^2 +b(x-x_0)^2 +c (y-y_0)^2},\\
\ \\
|x-x_0|, \ |y-y_0|, \ |t-t^{\pm}_c|\ll 1 ,
\ea
\eeq
where
\beq
\ba{l}
a=b_{12}\sin\phi_{1,0}(B+\sin\phi_{0,1})\sigma^2_{1,0}+\big[AB+\sqrt{(A^2-1)(B^2-1)}\\
-b_{12}\big(AB+\sqrt{(A^2-1)(B^2-1)}\big)\big]\sigma_{1,0}\sigma_{0,1}+b_{12}\sin\phi_{0,1}(A+\sin\phi_{1,0})\sigma^2_{0,1},\\
b=b_{12}\sin\phi_{1,0} k_1^2 B, \ \ c=b_{12}\sin\phi_{0,1} l_1^2 A.
\ea
\eeq

From these considerations one infers that, unless the NLS case, for which the amplitude of the $N$ breather solution has a fixed maximum achieved when the interference among its $N$ unstable modes is fully constructive, the amplitude of the AW $u_2$ can be arbitrarily large if the free parameters $t_{1,0},t_{0,1}$ are such that $t_{1,0}-t_{0,1}$ is sufficiently close to $\pm\Delta_c$. We expect similar and even richer features for the exact solutions \eqref{def_breather1}-\eqref{def_breather5_DS2} coming from the interaction of more than two unstable modes, and a subsequent paper will be devoted to a systematic study of these behaviors.

\subsection{Interesting limiting cases of the solution \eqref{xy_sol}-\eqref{xy_sol_param}}

\paragraph{The modes $\pm\vec k_{1,1},\pm\vec k_{1,-1}$ tend to the instability circle.} If $\pm\vec k_{1,1},\pm\vec k_{1,-1}$ tend to the instability circle: $k_1^2+l_1^2\to 4$, then $\phi_{1,0}+\phi_{0,1}\rightarrow\frac{\pi}{2}$ and $b_{12}\to\infty$. In addition the two growth rates become the same:
\beq
\sigma_{1,0}=\sigma_{0,1}=2\sin(2\phi)=:\sigma, \ \ \phi:=\phi_{1,0}.
\eeq
To perform this limit we choose $e^{\sigma_{1,0} t_{1,0}},e^{\sigma_{0,1} t_{0,1}} =O(b_{12})$, and we define the convenient $O(1)$ parameters
\beq
t_0=\frac{1}{\sigma}\log\left(\frac{e^{\sigma_{1,0} t_{1,0}}+e^{\sigma_{0,1} t_{0,1}}}{\sqrt{e^{2\sigma_{1,0} t_{1,0}}+e^{2\sigma_{0,1} t_{0,1}}}} \right), \ \xi =\arccos\left(\frac{e^{\sigma_{0,1} t_{0,1}}}{\sqrt{e^{2\sigma_{1,0} t_{1,0}}+e^{2\sigma_{0,1} t_{0,1}}}} \right),
\eeq
obtaining the following new solution in the limit
\begin{equation}
	u_b(x,y,t)= \frac{N_b(x,y,t)}{D_b(x,y,t)}e^{i\rho},
\end{equation}
where:
{\footnotesize
	\begin{equation}
		\begin{split}
			&N_{b}(x,y,t)=\cosh(\sigma (t-t_0) + 2 i \phi)-\sin\phi\sin\xi\cos(2\cos\phi\,(x-x_0))+\cos\phi\cos\xi\cos(2\sin\phi\,(y-y_0));
		\end{split}
	\end{equation}
}
and
{\footnotesize
	\begin{equation}
		\begin{split}
			&D_{b}(x,y,t)=\cosh(\sigma (t-t_0))+\sin\phi\sin\xi\cos(2\cos\phi\,(x-x_0))+\cos\phi\cos\xi\cos(2\sin\phi\,(y-y_0)).
		\end{split}
	\end{equation}
}
Since the two growth rates coincide in this limit, and we are left with only one time parameter $t_0$, the AW appearance consists of only one emergence. The blow up condition is achieved only if $\xi=\phi$.

\paragraph{Rational limit.} As in the NLS case, the rational limit to Peregrine like solutions is achieved taking the long wave limit in both $x$ and $y$ directions. Then we introduce the following notation
\begin{equation}
	k_1= \epsilon\,\delta_x, \hspace{1cm}	l_1= \epsilon\,\delta_y,  \hspace{1cm}\eps\ll 1,
\end{equation}
and, correspondingly, we choose the angles in the first quadrant
\beq
\phi_{1,0}=\frac{\pi}{2}-\eps \frac{\delta_x}{2}+O(\eps^3), \ \ \phi_{0,1}=\frac{\pi}{2}-\eps \frac{\delta_y}{2}+O(\eps^3),
\eeq
to get the nontrivial rational limit 
\begin{equation}
	u_{2P}(x,y,t)= \frac{N_P(x,y,t)}{D_P(x,y,t)}e^{i\rho},
\end{equation}
where
{\footnotesize
	\begin{equation}
		\begin{split}
			&N_P(x,y,t)=\\
			&\left(-3+4(x-x_0)^2+16(t-t_{1,0})(t-t_{1,0}-i)\right)\left(-3+4(y-y_0)^2+16(t-t_{0,1})(t-t_{0,1}+i)\right)+	\\
			&+64 i (t_{1,0}-t_{0,1})-128 (t-t_{1,0})(t-t_{0,1})-24,
		\end{split}
	\end{equation}
      }
      and
{\footnotesize
	\begin{equation}
		\begin{split}
			D_P(x,y,t)=&\left(-3+4(x-x_0)^2+16(t-t_{1,0})^2\right)\left(-3+4(y-y_0)^2+16(t-t_{0,1})^2\right)+	\\
			&+64 (t_{1,0}-t_{0,1})^2+16((x-x_0)^2+(y-y_0)^2).
		\end{split}
	\end{equation}
}

This $2+1$ dimensional generalization of the NLS Peregrine solution does not depend on the parameters $\delta_x$ and $\delta_y$. It describes the nonlinear interaction of two rational Peregrine walls of the type \eqref{Peregrine}, parallel to the $x$ and $y$ axes. Like the Peregrine solution, it  decreases rationally to the background \eqref{background} as $t\to\pm\infty$; it decreases to the orthogonal walls for $x^2+y^2\gg 1$, unlike the Peregrine solution that decreases to the background in this limit. As the solution \eqref{xy_sol}-\eqref{xy_sol_param}, it appears twice, but now the two appearances are described by the same function.  This rational solution, blowing up twice if $t_{1,0}=t_{0,1}$, in the space-time points $(x_0,y_0,t_{1,0}\pm \frac{\sqrt{3}}{4})$, should be a particular example of the general class of rational AWs solutions presented in \cite{Otha2}.   

\paragraph{Periodic-rational limit.}

If one performs the above long wave limit only, say, in the $y$ direction, the solution \eqref{xy_sol}-\eqref{xy_sol_param} leads to the following solution rational in y, periodic in x, and rational/hyperbolic in $t$: 
\begin{equation}
	u_{PR}(x,y,t)= \frac{N_{PR}(x,y,t)}{D_{PR}(x,y,t)}e^{i\rho},
\end{equation}
where
{\footnotesize
	\begin{equation}
		\begin{split}
			&N_{PR}(x,y,t)=\cosh(\sigma_{1,0}(t-t_{1,0})+2i\phi_{1,0})\left(-7+16(i+t-t_{0,1})(t-t_{0,1})+4(y-y_1)^2+4\csc^2\phi_{1,0}\right)\\
			&-8 \cot\phi_{1,0}\sinh(\sigma_{1,0}(t-t_{1,0})+2i\phi_{1,0})\left(i+2(t-t_{0,1})\right)+\\
			&+\sin\phi_{1,0}\cos(k_1(x-x_1))\left(-3 +16(i+t-t_{0,1})(t-t_{0,1})+4(y-y_1)^2\right),
		\end{split}
	\end{equation}
}
and
{\footnotesize
	\begin{equation}
		\begin{split}
			&D_{PR}(x,y,t)=\cosh(\sigma_{1,0}(t-t_{1,0}))\left(1+16(t-t_{0,1})^2+4(y-y_1)^2+4\cot^2\phi_{1,0}\right)\\
			&-16 \cot\phi_{1,0}\sinh(\sigma_{1,0}(t-t_{1,0}))\left(t-t_{0,1}\right)+\\
			&-\sin\phi_{1,0}\cos(k_1(x-x_1))\left(1 +16(t-t_{0,1})^2+4(y-y_1)^2\right).
		\end{split}
	\end{equation}
}
To the best of our knowledge this solution is new.

\section{Modulation instability and AW recurrence}

In this section we use the matched asymptotic expansions technique introduced in \cite{GS2} to describe the relevance of the above exact AW solutions in the DS2 doubly periodic Cauchy problem for AWs, in the case of one and two unstable modes.

The doubly periodic Cauchy problem for AWs of the focusing DS2 equation \eqref{DS}, $\eta=\nu=1$,  reads
\beq\label{Cauchy}
\ba{l}
u(x+L_x,y,t)=u(x,y+L_y,t)=u(x,y,t), \\ q(x+L_x,y,t)=q(x,y+L_y,t)=q(x,y,t), \\
u(x,y,0)=1+\eps \ v(x,y), \ \ q(x,y,0)=\eps \ w(x,y), \ \ 0< \eps\ll 1 ,
\ea
\eeq
where the initial perturbations can be expanded in Fourier modes as follows:
\beq
\ba{l}
v(x,y)=\sum\limits_{\mu,\nu\in\ZZ}c_{\mu,\nu}e^{i(k_{\mu}x+l_{\nu}y)},
\ea
\eeq

For $|t|\le O(1)$, the evolution is ruled by the linearized equations \eqref{linearized}, and the solution is described, through Fourier analysis and up to $O(\eps^2)$ terms, by the following formulas
\beq\label{sol_lin_1}
\ba{l}
u(x,y,t)=1+\eps\sum\limits_{
  m,n\in{\cal D}}
\Big(\frac{|\alpha_{m,n}|}{\sin(2\phi_{m,n})}\cos\left(k_m x+l_n y-\arg(\alpha_{m,n})-\pi/2 \right)e^{\Omega_{m,n} t+i\phi_{m,n}}\\
+\frac{|\beta_{m,n}|}{\sin(2\phi_{m,n})}\cos\left(k_m x+l_n y+\arg(\beta_{m,n})-\pi/2 \right)e^{-\Omega_{m,n} t-i\phi_{m,n}}\Big)+O(\eps)\mbox{-oscillations},
\ea
\eeq
\beq
\ba{l}
q(x,y,t)=\eps\sum\limits_{
  m,n\in{\cal D}}
\frac{\cos(2\theta_{m,n})}{\sin(\phi_{m,n})}\Big[|\alpha_{m,n}|\cos\left(k_m x+l_n y-\arg(\alpha_{m,n})-\pi/2 \right)e^{\Omega_{m,n} t}\\
  +|\beta_{m,n}|\cos\left(k_m x+l_n y+\arg(\beta_{m,n})-\pi/2 \right)e^{-\Omega_{m,n} t}\Big]+O(\eps)\mbox{-oscillations},
\ea
\eeq
where
\beq\label{def_alpha_beta}
\ba{l}
k_m=2\cos\phi_{m,n}\cos\theta_{m,n}, \ \ l_n=2\cos\phi_{m,n}\sin\theta_{m,n}, \\
\Rightarrow \ \ \ \phi_{m,n}=\arccos\left({\frac{\sqrt{k_m^2 +l_n^2}}{2}}\right), \ \
\theta_{m,n}=\arctan\left({\frac{l_n}{k_m}}\right), \\ 
\alpha_{m,n}=e^{-i\phi_{m,n}}\bar c_{m,n} -e^{i\phi_{m,n}}c_{-m,-n}, \\ \beta_{m,n}=e^{i\phi_{m,n}}\bar c_{-m,-n}-e^{-i\phi_{m,n}}c_{m,n},
\ea
\eeq
and

\beq
\mbox{
\footnotesize ${\cal D}=\left\{m\ge 1,~n\in\ZZ,~\left(\frac{m}{L_x} \right)^2+\left(\frac{n}{L_y} \right)^2<\frac{1}{\pi^2} \right\} \cup \left\{m=0, \ n\ge 1, \  \left(\frac{n}{L_y} \right)^2<\frac{1}{\pi^2} \right\}
$}.
\eeq

As time increases, the perturbation in \eqref{sol_lin_1} grows exponentially and, at $t=O(\log(1/\eps))$, it becomes order one and the dynamics is described by the fully nonlinear theory. It is when the exact solutions we constructed play a relevant role.

\subsection{One unstable mode}

In the case of one unstable mode we have the two cases \eqref{k10} and \eqref{k01}. \\
a) If $\vec k_{1,0}=(k_1,0)$ is the only unstable mode, i.e.:
\beq
\ba{l}
\pi< L_x<2\pi, \ L_y <\pi \ \ \ \ \Leftrightarrow \ \ \ \ 1< k_1 <2, \ \ l_1>2  \ \ \Leftrightarrow \\
\theta_{1,0}=0, \ k_1=2\cos\phi_{1,0}, \ 0<\phi_{1,0}<\pi/3 , \\
\Omega_{1,0}=k_1\sqrt{4-k_1^2}=2\sin(2 \phi_{1,0})=\sigma_{1,0},
\ea
\eeq
then equation \eqref{sol_lin_1} reduces to
\beq\label{lin_10}
\ba{l}
u(x,y,t)=1+\eps\Big[\frac{1}{\sin\left(2\phi_{1,0}\right)}\Big(|\alpha_{1,0}|\cos\left(2\cos\phi_{1,0} x-\arg(\alpha_{1,0})-\pi/2 \right)e^{i\phi_{1,0}+\sigma_{1,0}t}\\
+|\beta_{1,0}|\cos\left(2\cos\phi_{1,0} x+\arg(\beta_{1,0})-\pi/2 \right)e^{-i\phi_{1,0}-\sigma_{1,0}t}\Big)\Big]+O(\eps^2)\mbox{-oscillations}.
\ea
\eeq
Since the exact solution $u_{1,0}(x,y,t)$ in \eqref{u10} describes the nonlinear instability of the mode $\pm \vec k_{1,0}$, it is the natural candidate to describe the first AW appearance at $t=O\left(\log(1/\eps)\right)$. Then one chooses its appearance time $t^{(1)}$ as $t^{(1)}\equiv\frac{1}{\sigma_{1,0}}\log\frac{\gamma}{\eps}, \ \gamma>0$, with $\gamma$ to be fixed. In the intermediate time interval $1\ll t \ll O(\log(1/\eps))$, \eqref{lin_10} and $u_{1,0}(x,y,t)$ become
\beq\label{u_10_intermediate}
\ba{l}
u(x,y,t)\sim 1+\frac{\eps|\alpha_{1,0}|}{\sin\left(2\phi_{1,0}\right)}\cos\left[k_1 x-\arg(\alpha_{1,0})-\pi/2 \right]e^{i\phi_{1,0}+\sigma_{1,0}t}, \\
u_{1,0}(x,y,t)\sim e^{i\left(\rho^{(1)}-2\phi_{1,0}\right)}\left(1+\frac{2\eps}{\gamma}\sin(2\phi_{1,0})\cos\left(k_1(x-x^{(1)})\right)e^{\sigma_{1,0}t+i\psi_{1,0}} \right).
\ea
\eeq
Comparing the leading order asymptotics \eqref{u_10_intermediate} one fixes all the free parameters of $u_{1,0}$ as follows
\beq
\rho^{(1)}=2\phi_{1,0}, \ x^{(1)}=\frac{\arg(\alpha_{1,0})+\pi/2}{k_1}, \ t^{(1)}=\frac{1}{\sigma_{1,0}}\log\left(\frac{2\sin^2(2\phi_{1,0})}{\eps |\alpha_{1,0}|} \right),
\eeq
showing that the first AW appearance is described, to leading order and at $|t-t^{(1)}|=O(1)$, by
\beq\label{1st_appearance10}
u(x,y,t)=e^{2i\phi_{1,0}}{\cal A}(x-x^{(1)},t-t^{(1)},\phi_{1,0})+O(\eps),
\eeq
an elementary function of the initial data. We remark that, although the initial perturbation is an arbitrary doubly periodic function of $(x,y)$, since the only unstable mode is the horizontal mode $\pm\vec k_{1,0}$, the AW is one-dimensional and the $y$ dependence is confined at $O(\eps)$.

To describe analytically the AW recurrence, we also construct the first AW appearance at negative times, following the same strategy, obtaining
\beq\label{1st_appearance10_negative_t}
\ba{l}
u(x,y,t)=e^{-2i\phi_{1,0}}{\cal A}(x-x^{(0)},t-t^{(0)},\phi_{1,0})+O(\eps), \ \ |t-t^{(0)}|=O(1), \\
x^{(0)}=\frac{-\arg(\beta_{1,0})+\pi/2}{k_1}, \ \ \ t^{(0)}=-\frac{1}{\sigma_{1,0}}\log\left(\frac{2\sin^2(2\phi_{1,0})}{\eps |\beta_{1,0}|} \right).
\ea
\eeq
Then we compare the two consecutive appearances \eqref{1st_appearance10_negative_t} and \eqref{1st_appearance10}, and using the time translation property of the model, we infer that the dynamics is described by a FPUT recurrence of AWs, and that the $j^{th}$ appearance is described by 
\beq\label{jth_appearance10}
u(x,y,t)=e^{i\rho^{(j)}}{\cal A}(x-x^{(j)}_{1,0},t-t^{(j)}_{1,0},\phi_{1,0})+O(\eps), \ \ |t-t^{(j)}_{1,0}|=O(1), \ \ j\ge 1,
\eeq
where
\beq
\ba{l}
\rho^{(j)}_{1,0}=\rho^{(1)}_{1,0}+(j-1)4\phi_{1,0}, \ \ x^{(j)}_{1,0}=x^{(1)}_{1,0}+(j-1)\frac{\arg(\alpha_{1,0}\beta_{1,0})}{k_1}, \\
t^{(j)}_{1,0}=t^{(1)}_{1,0}+(j-1)\frac{2}{\sigma_{1,0}}\log\left(\frac{2\sin^2(2\phi_{1,0})}{\eps \sqrt{|\alpha_{1,0}\beta_{1,0}|}} \right).
\ea
\eeq
\ \\
b) If $\vec k_{0,1}=(0,l_1)$ is the only unstable mode, i.e.:
\beq
\ba{l}
\pi< L_y<2\pi, \ L_x <\pi \ \ \ \ \Leftrightarrow \ \ \ \ 1< l_1 <2, \ \ k_1>2  \ \ \Leftrightarrow \\
\theta_{1,0}=\pi/2, \ \theta_{0,1}=0, \ l_1=2\cos\phi_{0,1}, \ 0<\phi_{0,1}<\pi/3 , \\
\sigma_{0,1}=l_1\sqrt{4-l_1^2}=2\sin(2 \phi_{0,1})=-\Omega_{0,1},
\ea
\eeq
now the nonlinear stages of MI are described, to leading order, by the exact solution $u_{0,1}$ and, proceeding as before, one can show that the solution of the Cauchy problem \eqref{Cauchy} is described by a FPUT recurrence of AWs, and that the $j^{th}$ appearance is described by 
\beq\label{jth_appearance01}
u(x,y,t)=e^{i\rho^{(j)}_{0,1}}{\cal A}(y-y^{(j)}_{0,1},t-t^{(j)}_{0,1},\phi_{0,1})+O(\eps), \ \ |t-t^{(j)}_{0,1}|=O(1), \ \ j\ge 1,
\eeq
where
\beq
\ba{l}
\rho^{(j)}_{0,1}=\rho^{(1)}_{0,1}-(j-1)4\phi_{0,1}, \ \ y^{(j)}_{0,1}=y^{(1)}_{0,1}-(j-1)\frac{\arg(\alpha_{0,1}\beta_{0,1})}{l_1}, \\
t^{(j)}_{0,1}=t^{(1)}_{0,1}+(j-1)\frac{2}{\sigma_{0,1}}\log\left(\frac{2\sin^2(2\phi_{0,1})}{\eps \sqrt{|\alpha_{0,1}\beta_{0,1}|}} \right),
\ea
\eeq
and
\beq
\rho^{(1)}_{0,1}=-2\phi_{0,1}, \ y^{(1)}_{0,1}=\frac{-\arg(\beta_{0,1})+\pi/2}{l_1}, \ t^{(1)}_{0,1}=\frac{1}{\sigma_{0,1}}\log\left(\frac{2\sin^2(2\phi_{0,1})}{\eps \sqrt{|\beta_{0,1}|}}\right).
\eeq

We remark that, in both cases, since we have only one growing mode (horizontal or vertical) in the overlapping region, and since the Akhmediev type solution, describing the growth of this unstable mode, contains enough free parameters for a successful matching, the remaining mismatch cannot affect the leading order behavior at the appearance. Therefore this stability argument plus uniqueness of the DS2 evolution imply that the appearance of the AW is described by the one dimensional Akhmediev solution, and the dependence on both $x$ and $y$ variables is hidden at $O(\eps)$.

\subsection{Two unstable modes}

The simplest truly two dimensional AW dynamics takes place when there are only the two unstable modes $\pm\vec k_{1,0}$ and $\pm\vec k_{0,1}$ (see the bottom left picture of Figure \ref{instDS2b}), corresponding to the constraint
\beq\label{constr}
\ba{l}
\pi<L_x,L_y <2\pi, \ \ \frac{1}{L_x^2}+\frac{1}{L_y^2}>\frac{1}{\pi^2}, \ \Leftrightarrow \\
1<k_1,l_1<2, \ \ \ \ k_1^2 +l_1^2>4 ,  \ \Leftrightarrow \\
0<\phi_{1,0},\phi_{0,1}<\pi/3, \ \ \cos^2\phi_{1,0}+\cos^2\phi_{0,1}>1 .
\ea
\eeq

Then the linear stage of MI \eqref{sol_lin_1}, for $|t|\le O(1)$, reduces to
\beq\label{lin_10_01}
\ba{l}
u(x,y,t)=1+\eps\Big[\frac{1}{\sin\left(2\phi_{1,0}\right)}\Big(|\alpha_{1,0}|\cos\left(2\cos\phi_{1,0} x-\arg(\alpha_{1,0})-\pi/2 \right)e^{i\phi_{1,0}+\sigma_{1,0}t}\\
+|\beta_{1,0}|\cos\left(2\cos\phi_{1,0} x+\arg(\beta_{1,0})-\pi/2 \right)e^{-i\phi_{1,0}-\sigma_{1,0}t}\Big)\\
+\frac{1}{\sin\left(2\phi_{0,1}\right)}\Big[|\alpha_{0,1}|\cos\left(l_1 y-\arg(\alpha_{0,1})-\pi/2 \right)e^{-\sigma_{0,1}t+i\phi_{0,1}}\\
+|\beta_{0,1}|\cos\left(l_1 y+\arg(\beta_{1,0})-\pi/2 \right)e^{\sigma_{0,1}t-i\phi_{0,1}}\Big]
+O(\eps)\mbox{-oscillations}. 
\ea
\eeq

Reasoning as before, since the exact solution $u_2(x,y,t)$ of DS2 in \eqref{xy_sol}-\eqref{xy_sol_param} describes the nonlinear interaction of the unstable modes $\pm \vec k_{1,0}$, $\pm \vec k_{0,1}$, it is the natural candidate to characterize this nonlinear stage, and following exactly the same strategy as before, we find that the first AW appearance is described to leading order by the solution \eqref{xy_sol}
\beq\label{xy_1st_appearance}
u(x,y,t)=u_2(x,y,t;\phi_{1,0},\phi_{0,1},x^{(1)},y^{(1)},t^{(1)}_{1,0},t^{(1)}_{0,1},\rho^{(1)})+O(\eps),
\eeq
where the solution parameters are expressed in terms of the initial data as follows
\beq\label{xy_1st_appearance_parameters}
  \ba{l}
  \rho^{(1)}=2\left(\phi_{1,0}-\phi_{0,1} \right), \ x^{(1)}=\frac{\arg(\alpha_{1,0})+\pi/2}{k_1}, \ y^{(1)}=\frac{-\arg(\beta_{0,1})+\pi/2}{l_1}, \\
 t^{(1)}_{1,0}=\frac{1}{\sigma_{1,0}}\log\left(\frac{2 b_{12}\sin^2\left(2\phi_{1,0}\right)}{\eps |\alpha_{1,0}|}\right), \ t^{(1)}_{0,1}=\frac{1}{\sigma_{0,1}}\log\left(\frac{2 b_{12}\sin^2\left(2\phi_{0,1}\right)}{\eps |\beta_{0,1}|}\right). 
\ea
\eeq

Therefore the first appearance of the AW in the Cauchy problem consists of the two emergences described by the exact solution \eqref{xy_sol}-\eqref{xy_sol_param} (see Figure \ref{AW2}), whose parameters are expressed in terms of the initial data through elementary functions.

As for the case of one unstable mode, we remark that, since we have only two growing modes in the overlapping time region, and since the exact solution \eqref{xy_sol}-\eqref{xy_sol_param}, describing the growth and the nonlinear interaction of these unstable modes, contains enough free parameters for a successful matching, the remaining mismatch cannot affect the leading order behavior. Therefore this stability argument plus uniqueness of the DS2 evolution imply that the first appearance of the AW is described by the solution \eqref{xy_1st_appearance},\eqref{xy_1st_appearance_parameters}, an elementary function of the initial data.

\begin{figure}[H]
  \centering
  \includegraphics[width=6.5cm,height=4cm]{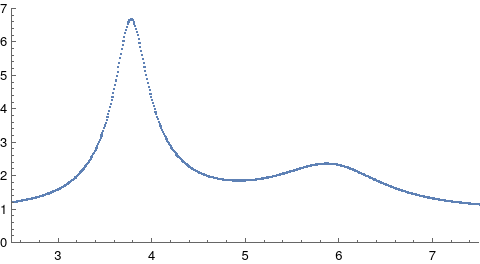} \ 
	\includegraphics[width=6.5cm,height=4cm]{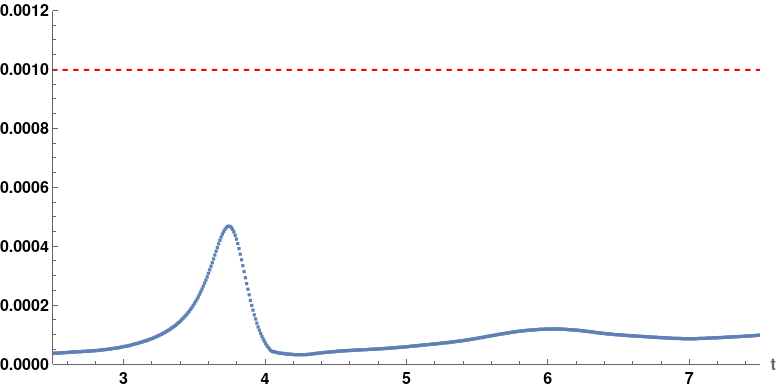}
	\caption{Here we study the two emergences of AWs in the time interval of the first appearance, for the initial data $\eps=10^{-3}$, $c_{1,0}=0.8+i0.4$, $c_{-1,0}=1.2-i0.1$, $c_{0,1}=-0.64-i0.3$, $c_{0,-1}=0.5+i0.2$. Left picture: the max of the amplitude of the AW $\| u_{num}\|_{\infty}(t)$ as function of time, where $u_{num}$ is the numerical solution; the first emergence at $(x,y,t)=(3.24019,1.227442,3.780)$ with a peak of height $6.6786$; the second emergence at $(x,y,t)=(3.24019,1.227442,5.868)$ with a peak of smaller height $2.3631$. Right picture: the uniform distance $\| u_{num}-u_2\|_{\infty}(t)$ between the analytic solution $u_2$ \eqref{xy_1st_appearance},\eqref{xy_1st_appearance_parameters} and the numerical solution $u_{num}$; the two peaks of the distance correspond exactly to the two AW emergences of the left picture, and the distance remains always $\le 5\cdot 10^{-4}$, smaller than the estimated error from theoretical considerations $O(10^{-3})$, indicated by the horizontal dotted line.}\label{distance}
\end{figure}

To have an idea of how well the analytic solution $u_2$ in \eqref{xy_1st_appearance},\eqref{xy_1st_appearance_parameters} describe the first appearance of the AW in the AW Cauchy problem, we evaluate the uniform distance between $u_2$ and the numerical solution $u_{num}$ (obtained using the $4^{th}$ order split step Fourier method \cite{SSFM}):
\beq
\| u_{num}-u_2\|_{\infty}(t):=\sup_{x\in [0,L_x], y\in [0,L_y]}|u_{num}(x,y,t)-u_{2}(x,y,t)| ,
\eeq
in the time interval in which the AW first appears, see Figure \ref{distance}. The agreement is excellent, since the error is much smaller than expected from theoretical considerations.

We end this paper with some considerations on the possibility of blow up in the first appearance of the AW \eqref{xy_1st_appearance},\eqref{xy_1st_appearance_parameters}. If we compare the difference between the time parameters $t^{(1)}_{1,0}-t^{(1)}_{0,1}$ coming from the Cauchy problem
\beq
t^{(1)}_{1,0}-t^{(1)}_{0,1}=\frac{1}{\sigma_{1,0}}\log\left(\frac{2 b_{12}\sin^2\left(2\phi_{1,0}\right)}{\eps |\alpha_{1,0}|}\right)-\frac{1}{\sigma_{0,1}}\log\left(\frac{2 b_{12}\sin^2\left(2\phi_{0,1}\right)}{\eps |\beta_{0,1}|}\right)
\eeq
with the critical difference \eqref{critical_parameter},\eqref{def_A_B} corresponding to the blow up of the solution \eqref{xy_sol}, we infer that we have blow up if one of the following two equations are satisfied
\beq\label{blowup-}
\left(\frac{b_{12}\sigma^2_{1,0}\left(B+\sqrt{B^2 -1} \right)}{2\eps |\alpha_{1,0}|} \right)^{\sigma_{0,1}}=\left(\frac{b_{12}\sigma^2_{0,1}\left(A+\sqrt{A^2 -1} \right)}{2\eps |\beta_{0,1}|}  \right)^{\sigma_{1,0}},
\eeq
\beq\label{blowup+}
\left(\frac{b_{12}\sigma^2_{1,0}}{2\eps |\alpha_{1,0}|\left(B+\sqrt{B^2 -1} \right)} \right)^{\sigma_{0,1}}=\left(\frac{b_{12}\sigma^2_{0,1}}{2\eps |\beta_{0,1}|\left(A+\sqrt{A^2 -1} \right)}  \right)^{\sigma_{1,0}}.
\eeq
If \eqref{blowup-} holds, then blow up occurs at the first emergence; if \eqref{blowup+} holds, then blow up occurs at the second emergence.

\begin{figure}[H]
	\centering
	\includegraphics[width=6.5cm,height=6.5cm]{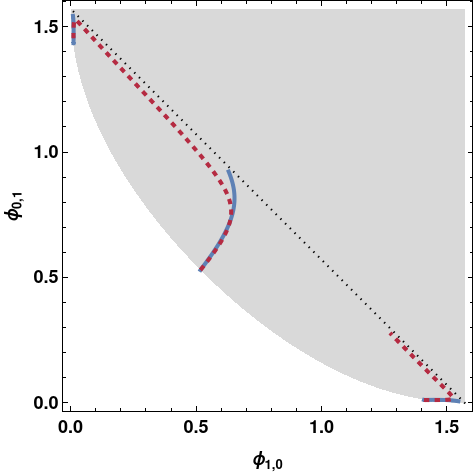}\includegraphics[width=6.5cm,height=6.5cm]{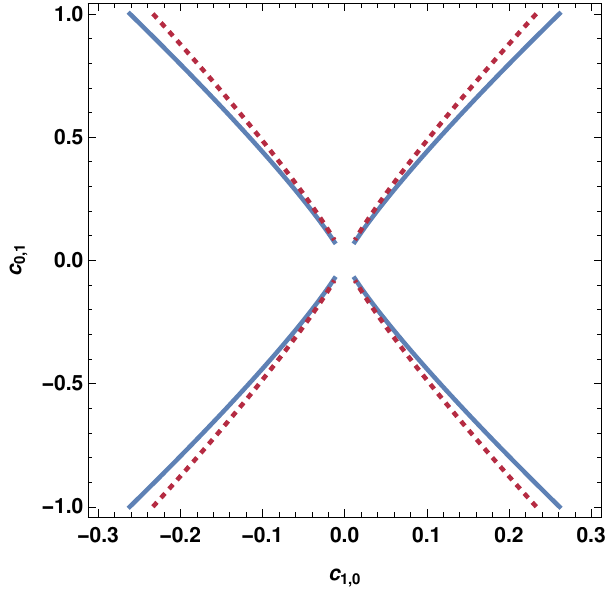}
	\caption{The left picture shows the solid and dashed curves in the $(\phi_{1,0} ,\phi_{0,1})$ plane on which the blow-up conditions \eqref{blowup-} and respectively \eqref{blowup+} are satisfied, for the initial data  $c_{1,0}=0.8+i0.4$, $c_{-1,0}=1.2-i0.1$, $c_{0,1}=-0.64-i0.3$, $c_{0,-1}=0.5+i0.2$ and $\epsilon=10^{-3}$. The gray area ($\sin\phi_{1,0}+\sin\phi_{0,1}>1$) is the region where blow-up can occour, and the straight dotted line  $\phi_{1,0}+\phi_{0,1}=\frac{\pi}{2}$ separates the gray region in two parts, the region below, where only two modes are unstable, and the region above, where more than two modes are unstable. The right picture shows the solid and dashed curves in the $(c_{1,0},c_{0,1})$ plane on which the blow-up conditions \eqref{blowup-} and respectively \eqref{blowup+} are satisfied, for $\phi_{1,0}=0.7$, $\phi_{0,1}=0.5$, and $\epsilon=10^{-3}$. }\label{blowup_plane}
\end{figure}

Equations \eqref{blowup-},\eqref{blowup+} depend on the initial data parameters $\eps,c_{m,n}$, and on the unstable mode parameters $\phi_{1,0},\phi_{0,1}$. If, for instance, we fix the initial condition parameters, then the blow up regions in the $(\phi_{1,0},\phi_{0,1})$ plane are curves (see the left picture in Figure \ref{blowup_plane}). If we fix instead the unstable mode parameters $(\phi_{1,0},\phi_{0,1})$, in the space of real initial data of the type
\beq
u(x,y,0)=1+2\eps \left[ c_{1,0}\cos\left(2\cos\phi_{1,0}x\right)+c_{0,1}\cos\left(2\cos\phi_{0,1}y\right)\right], \ \ c_{1,0},c_{0,1}\in\RR,
\eeq
the blow up regions in the $(c_{1,0},c_{0,1})$ plane are again curves (see the right picture  in  Figure  \ref{blowup_plane}).
Since curves have zero measure in the plane, we conclude that, generically, the first appearance of the AW does not give rise to blow up. But the amplitude of the AW can be arbitrarily large if the parameters are sufficiently close to the singular curves, and situations of this type are expected to take place at later times, during the recurrence. 

Similar considerations are expected to be valid for a generic Cauchy problem for AWs involving more than two unstable modes, and will be the subject of future investigation.

\section{Conclusions and future perspectives}

In this paper we investigated MI and AWs of the integrable DS1 and DS2 equations. More precisely, 1) we constructed the $N$-breather AW solution of Akhmediev type of the integrable DS1 and DS2 equations. 2) We selected, in the DS2 case, the subclass of AW solutions that are relevant in the Cauchy problem for periodic AWs, and, in the case of the simplest multidimensional solution, i) we identified the constraint on its arbitrary parameters giving rise to blow up at finite time; ii) we constructed its limiting cases. 3) We used matched asymptotic expansions to describe the relevance of the constructed AW exact solutions in the DS2 doubly periodic Cauchy problem for AWs, in the case of one and two unstable modes, showing in particular that blow up is not generic.

This paper opens several research directions we plan to follow in the near future. 1) The proper implementation of the finite gap formalism developed in \cite{GS5} to solve, in terms of elementary functions of the generic initial data, the general periodic Cauchy problem of AWs for the DS2 equation in the case of a finite number $N>1$ of unstable nonlinear modes, since matched asymptotic expansions are not adequate to study AW recurrence in this case.   2) The use of the analytic solution of the AW Cauchy probòem to study the probability of generating multidimensional AWs of amplitude greater than a certain critical value in a given time interval. 3) The generalization to multidimensions of the perturbation theory of AWs developed for 1+1 dimensional NLS type equations in \cite{Coppini1,Coppini2,Coppini3,CS_AL2}, to describe analytically the order one effects of physical perturbations of the DS2 equation on the AW dynamics.    

\vskip 10pt
\noindent
{\bf Acknowledgments}. The work of F. Coppini and  P. M. Santini was supported by the Research Project of National Interest (PRIN) No. 2020X4T57A. It was also done within the activities of the INDAM-GNFM. The work of P. G. Grinevich was supported by the Russian Science Foundation under grant no. 21-11-00331, https://rscf.ru/en/project/21-11-00331/ .

\end{document}